# A method for finding distribution of metabolic energy between organismal functions: application to birds' energy expenditures to counteract gravity and to support steady and short flights

Yuri K. Shestopaloff

*Consultant, Prof. Dr. Sci. (Phys.)*
*Toronto, Ontario, Canada*
*shes169@yahoo.ca*

**Abstract**

Production of energy (metabolism) and its distribution is vital for living organisms, both at individual level - between different functions of an organism, as well as between species of communities at different organizational levels, including food chains. Here, a new general method for finding distribution of metabolic energy between different organismal functions is proposed. The method is based on earlier discoveries (in two independent studies, for multicellular and unicellular organisms) that metabolic allometric scaling is the result of natural selection guided by optimization of distribution of common resources between the species of a food chain. This distribution is established in such a way that it secures amount of resources for each species to reproduce in sufficient quantities (sufficient for preservation of a food chain), while not allowing some species acquiring too many resources to jeopardize existence of other species they prey on or share common resources with. The introduced method was applied to birds, including both steady and short flights modes. Birds' metabolism has specifics, because besides other functions, it has to compensate force of gravity during the flight. However, this parameter is difficult to find, while such data is important for ecological studies at population and individual levels. We discovered increase of fraction of metabolic power required to compensate force of gravity with mass increase, which is a principal factor, restricting maximum possible mass of flying animals. The obtained results show efficiency of a proposed method for detailed studies of animals' metabolism, physiology, and of population ecology.

*Keywords*: flying animals; metabolism; maximum mass; energy expenditures; energy distribution; force of gravity; acceleration; potential energy; true take off

# 1. Introduction

Energy production is a very foundation of life as such. In living organisms, the rate of produced energy ($E$) is measured as amount of energy produced per unit of time (J s$^{-1}$), and sometimes, in addition, per unit of mass (J s$^{-1}$ kg$^{-1}$). In many instances, dependence of rate of produced energy from mass of an organism $M$ is described mathematically as a power equation.

$$E = aM^{b_x} \quad (1)$$

Here, the power $b_x$ is called a *metabolic allometric exponent*. In most instances, it is less than one, in which case the rate of energy production increases *slower* than mass. The regularity of change of energy production expressed by Eqn. 1 puzzled scientists from the day of its discovery by Kleiber in 1932, who was the first to calculate the value of $b_x = 0.74$ across *different* species. (In this case, the value presents *interspecific* metabolic allometric scaling, which we consider in this paper. When such scaling is considered for the same species, or ontogenetically, it is called *intraspecific* allometric scaling. The fundamental reasons of both phenomena, despite the outer similarity and resemblance of terminology, are quite different, as it is discussed in (Shestopaloff, 2016a).) Many explanations of interspecific metabolic allometric scaling, often contradicting each other, were proposed since then. The present consensus among most researchers is as follows:

(a) Metabolic allometric scaling is due to cooperative action of different factors, shaping organismal metabolism through mechanisms of natural selection;

(b) The value of metabolic allometric exponent, when applied to certain subsets of animals, is defined by different combinations of different factors, so that there is no even a single deterministic base, which could be composed of some key mechanisms, for this phenomenon.

Thus, within the last two decades the most productive studies of animals' metabolism, either implicitly or explicitly, were framed by these concepts. The significance of such studies for ecological purposes, like ecological life histories, is well understood (Witting, 2024). In recent years, greater attention was given to the aspect of animal metabolism related to the role of natural selection in optimization and adaptation of species' metabolism to the environments they live in. From ecological perspective, metabolism is of importance for not only individual living organisms, but it also plays an essential role in functioning of their communities. Such, the authors of work (Szangolies et al., 2024), investigating a metabolic community model for a terrestrial mammal community, conclude: "We argue that energetics should be part of community ecology theory, as the relative energetic status and reproductive investment can reveal why and under what environmental conditions coexistence is likely to occur."

We will use results of two independent studies (Shestopaloff, 2024a, 2024b), for multicellular and unicellular organisms accordingly. (Note that the first versions of these papers (Shestopaloff, 2016b, 2016c) were made publicly available in 2016, which eventually led to the aforementioned 2024 publications, when in recent years some original ideas, such as the one about the role of natural selection, began taking grounds in research community.) In these two separate studies, it was independently discovered and proved that the value of a metabolic allometric exponent is a quantitative reflection of an optimal distribution of resources between the species of a food chain populating some habitat. This optimal distribution is a principal factor providing dynamic stability of food chains, which is achieved through natural selection, guided by this evolutionary principle. Thus attained optimal distribution of resources within the food chain secures reproduction of its species in quantities, sufficient for continuous preservation and integrity of the food chain. At the same time, thus established equilibrium caps amount of resources available for each species, so that no species could obtain too many resources, and through that acquire too strong competitive advantage resulting in their excessive reproduction, which in turn can jeopardize the integrity of a food chain through suppression or even extinction of other species in the food chain. In the course of such a balanced distribution of resources, food chain preserves its dynamic continuity. When this balanced distribution is broken, the food chain is destroyed, and a new food chain will evolutionarily develop, whose creation will be guided by the same principle of optimized distribution of nutrients between the species of a food chain. If one thinks for a moment, if not for such a balance in nutrients distribution, the food chain's more or less prolonged existence would be *in principle* impossible, since otherwise the only alternative is chaos and disorder. However, this is not what we observe in nature. (This is rather the attribute of modern human societies, whose governance ignores such *natural* balancing principles, and so are the results of this governance.)

This fundamental principle of life organization of different species sharing resources of the same habitat is rather a universal one. The aforementioned study (Shestopaloff, 2016c, 2024a) considered *multicellular* organisms, in particular mammals, reptiles, fish and birds. However, exactly the same organizational evolutionary principle was independently discovered for *unicellular* organisms in (Shestopaloff, 2016b, 2024b), when each next unicellular organism with bigger mass is having some advantage, acquiring more nutrients per unit surface from the same environment, but not an excessive one, so that the entire food chain of microorganisms can sustain.

In this paper, we propose a method for finding energy distribution for supporting different organismal functions in animals in general, and consider from this perspective metabolism of birds, and in particular energy distribution for supporting steady cruising in the air, like in long distance travels, providing energy at rest (basal metabolic rate, BMR), making frequent take-offs at positive angles in short flights (also called true take-offs), compensating the force of gravity, and acquiring potential energy during ascends. First, we will consider steady flights, which do not require spending energy for acceleration at upward take-offs, and energy for acquisition of potential energy during ascends. Then, we consider short flights, which besides spending energy for counteracting gravity and supporting steady flight speed also include two last features.

In this study, compensation of force of gravity presents especial interest, for several reasons. One is its possible practical value for humans' flight related activities, such as energy optimization, and also regarding principal factors imposing weight limitations on flying animals. Besides, energy production and distribution closely relate to nutrient consumption, which is in turn tightly coupled with many ecological issues. There are studies, considering different components of birds' energy budget, such as basal, perching, hovering, flying, torpor metabolic rates, thermoregulatory costs (Shankar et al., 2020). However, none of them could evaluate how much energy birds have to spend to counteract force of gravity, while this is a significant in terms of spent energy and always presenting flight factor. Discovering how particularly natural selection and adaptation worked to incorporate such a very special factor as force of gravity during flight could be instructive from a scientific point of view, and also could shed light on many aspects of still evolving theory of natural selection.

**2. Materials and Methods**

This paper is presenting *method* for studying composition of metabolic energy expenditures based on *indirect* measurements of energy components, namely through functional characteristics, such as animals' speed, which these energy components support. The case of birds in this regard is of interest because there are at least two distinct functions the birds have to support during flights - one is to provide a certain speed of motion in the air, and the other is compensation of gravity. Certainly, both functions interfere in some way, but still, the need for gravity compensation in inescapable.

**2.1. Methodological basis of metabolic studies**

The main vehicle in studies of animal metabolism is measurement of produced energy through oxygen consumption, which can be added with other means, such as, for instance, measuring heat dissipated by animals' bodies. Accounting only for oxygen consumption excludes from consideration anaerobic energy producing mechanisms, which also may contribute to energy production, especially at high levels of physical activity. This could be the case with maximal metabolic rate in athletic animals, for whose subset a higher value of allometric exponent was obtained by indirect method, than the one accessed by oxygen consumption only (0.985 in (Shestopaloff, 2016c, 2024a) versus 0.942 in (Weibel and Hoppeler, 2005)).

As most general problems, metabolic studies include set of metabolic allometric theories, attempting to explain experimental results. The most known relate to certain physiological constraints, associated with physical restrictions, heat dissipation needs and limitations, nutrient influx and waste removal in the context of volume-surface relationship, and natural selection (Witting, 2024). With the advent of Big Data era, a newer trend making deductions based on analysis of large volumes of data without prior hypotheses might look tempting, but fortunately it already showed its fallacy in other scientific disciplines, of which societal studies produced probably the most vivid and convincing illustrations (Turchin, 2023). Scientific hypotheses arising on the basis of experiments and other inputs, is what makes the spirit and essence of science.

## 2.2. Methodological improvements of the presented study

The present study, by and large, suggests adding to the described above experimental and theoretical toolset of metabolic ecological studies *two more tools*.

(1) The first one is a general concept, namely the one that the metabolic allometric scaling is the result of natural selection, guided by optimization of distribution of resources of a certain habitat between the species of a food chain of this habitat. (The author thinks that this principal is applied at other organizational levels as well.) This could be not a small enhancement of the methodological capabilities of studies of animal metabolism, which will provide a more robust and definitive framework, as well as several invariant characteristics, to which metabolic properties of different organisms can be related.

(2) The second and the main thread of the paper is a more particular method of indirect determination of energy components supporting different organismal functions. This problem is of importance for many ecological issues associated with energy production and consumption, both at population and organismal levels. For that, we use an example of energy distribution in birds to support different modes of flight, and the energy required to compensate

*Yuri K. Shestopaloff*

the force of gravity, as well as the energy supporting the basal metabolic rate. The energy required to counteract gravity, in turn, can be used for evaluation of maximal mass of flying animals.

**2.3. Metabolic allometric scaling in birds**

To introduce and validate the proposed method, we need data on metabolic allometric exponents in birds, and birds' speeds. An extensive review (Swanson and Garland, 2009) of measured avian metabolic allometric exponents (with emphasis on summit metabolism and cold tolerance, but not only) concludes: "Reported values ... vary, but are typically in the range of 0.65-0.75." Variations occur across different division lines, such as origins of populations used for metabolic measurements, captive-raised or wild-caught, diet-correlated changes, passerine or nonpasserine, different sex - male versus female, desert versus non-desert ones, old desert or newer desert species, etc. Sometimes the studies aimed at the same composition of species produce conflicting results. Apparently, the used experimental methodology also can contribute to diversions.

Authors also note that "maximum flight metabolic rates generally exceed BMR by 8 to 14-fold for flying birds". Two values of allometric exponent for flights were presented: 0.804 and 0.813.

Another, probably the largest of its kind, study (McNab, 2009) confirmed substantial variations and dependence of allometric exponents on the combinations of factors included into consideration. The main propositions the author sets forth are as follows.

(a) All advocates of a universal power for metabolism ignore its residual variation, as if it were simply experimental error or inconvenient. A close examination, however, demonstrates some fundamental differences in the energetics of birds and mammals, one of which is that birds collectively have BMRs that are much higher than those of mammals of the same mass.

(b) Birds have BMRs that are approximately ... 2.1 times those of mammals (at a mass of 1 g). But because the scaling power in birds is lower (0.652) than that of mammals (0.721), the difference in BMR between birds and mammals decreases with an increase in mass: these two equations reach equality at a mass of 38.3 kg.

The last observation and the mass value are worth noting. This evaluation of mass corresponds to about the largest known flying birds. The result can be interpreted in such a way that the metabolic biochemical machinery becomes rather impractical (or incapable) for producing so much energy for supporting flight of heavier animals. Our results presented below actually confirm this assumption, explaining which factor is responsible for such an arrangement of metabolic biochemical machinery imposing a constraint on the maximal mass of flying animals.

Given the surveyed variations of metabolic allometric exponents for birds in general, which is excessive for our purposes, we will consider a more restrictive set of birds with more definitive values of metabolic allometric exponents from works (Nudds and Bryant, 2000; Ronning et al, 2005), adding previously mentioned values for steady flight, for which these works provide few data. We will use values: $b_{BMR} = 0.662$ for BMR, $b_{SF} = 0.802$ for a steady flight, obtained as the average of three known to us values; $b_{ShF} = 0.874$ for short flight. These values are commensurate with data from other sources. Such, the allometric exponent 0.874 for short flight is close to the adjusted scaling of 0.88 for the maximal metabolic rate for exercising birds from (Glazier, 2005). Close in value allometric exponents for BMR are also available in the literature.

We need to elaborate on the notion of "short flight", since this is an important for our undertakings flight mode, in which birds exercise maximum metabolic power. In short flights, both distances and flight time are short, birds frequently exercise true take-offs, landings, ascents, descents and maneuvering at speeds below minimum power speed (Nudds and Bryant, 2000). In the same work, the authors measured energy expenditures of 13 g zebra finches *Taeniopygia guttata* engaging in repeated short flights. Compared to 'nonflying' controls, short flight energy expenditures were 27.8 times of zebra finches BMR, which is over three times the predicted flight expenditure derived from existing aerodynamic models. Since the aerodynamic models can estimate only energy expenditures for a steady flapping flight, despite the approximate nature of aerodynamic models', this still gives an idea how greater can be energy expenditures required to support short flights, which makes our choice of short flights as the most energy demanding mode more justified.

**2.4. Energy components of short flight**

From the physical point of view, we can distinguish the following distinct operations a bird frequently performs during short flights:

(a)  Work against force of gravity to acquire potential energy $A$ during the ascent, which is $A = Mgh$ (2).

Here $M$ is mass, $h$ is height of an ascend; $g$ is a gravitational constant, free fall acceleration, $g = 9.81 \, \text{m s}^{-2}$.

(b)  Work $K$ required accelerating the bird's body at take-off, from zero to speed $v$, thus acquiring kinetic energy $K = Mv^2 / 2$ (3).

Such a simplistic approach to evaluation of kinetic energy of a flying bird can be questioned at a finer level of details, but for our purpose - just to get an idea of the order of magnitude of required kinetic energy - using Eqn. 3 is sufficient.

(c) Energy required to support steady flight. One of its components, for instance, is the energy needed to overcome the air resistance.

Let us get an idea of the value of potential energy versus the energy needed for acceleration using Eqns. 2 and 3. Suppose a bird takes-off at a positive angle up to a height of $h$=10 m, reaching at this height speed 10 m s$^{-1}$. Then we have

$$A/K = Mgh/(Mv^2/2) = 2gh/v^2 = 2 \times 9.81 \times 10/10^2 \approx 1.96$$

In other words, energy required for acceleration is of the order of magnitude of energy needed to acquire potential energy during regular short flights.

There is also energy needed to compensate force of gravity. However, this one is difficult to calculate using only physical approach. For instance, we can hold some weight by a stretched forward arm. Obviously, some *biochemical* energy is spent for that. We can measure the heart rate and see that it was increased, which is the proof of increased energy expenditure. However, we have no distance, no speed to calculate this energy using *physical* laws.

Now, we will introduce the mathematical apparatus, which will help us to decompose the total metabolic energy into its constituents.

## 2.5. Constituents of a metabolic allometric exponent

While the values of metabolic allometric exponents for mammals, reptiles and fish, found in (Shestopaloff, 2016c, 2024a), correspond to experimental values well, the obtained by the author value of allometric exponent for the maximal metabolic rate of 0.771 for birds was noticeably less than experimental values in the range of 0.84-0.88, reported in the literature. On the other hand, the calculated values of exponents for the basal metabolic rate of birds matched results of experimental studies very well. Such a fact led to an assumption that there are some additional factors, presenting during high birds' activity, which is flight, affecting their maximal metabolic rate. Analysis of possible factors showed that such an unaccounted factor could be the action of force of gravity and some other factors. Here, we consider in detail, how the force of gravity can be included into calculation of exponent for the maximal metabolic rate in birds, and, based on this approach, find energy characteristics, required to compensate force of gravity in a *steady flight* first, and then how much energy is needed to support acceleration and change of potential energy in energetically more demanding *short flights*.

It was derived in (Shestopaloff, 2016c, 2024a) that the allometric exponent for birds can be calculated using formula (24) from the main text of that paper, which is as follows.

$$b_x = (1 - b_l)b_{sk} + b_v \quad (4)$$

Here, $b_l$, $b_{sk}$, $b_v$ accordingly denote allometric exponents for the wingspan, skeleton mass, and speed of birds.

It was found in Appendix D.3 in (Shestopaloff, 2024a) that $b_l = 0.399 \pm 0.031$, $b_{sk} = 1.073$. The value of $b_v$ is the parameter we need to find. Let us calculate the value of the first term:

$$b_f = (1 - b_l)b_{sk} = (1 - 0.399) \times 1.073 = 0.645 \quad (5)$$

Since the force of gravity is omnipresent, it affects all birds with the same acceleration of free fall *g*. The fundamental biochemical mechanisms providing energy in multicellular living species generally do not differ much, so as the first approximation we may assume the same amount of generated energy per unit of resources (like per kilo of food). Based on these assumptions, we may assume that the energy required to counteract the force of gravity $F_L = Mg$ is proportional to mass *M*. On the other hand, metabolic power *E* increases slower than mass, that is $E = aM^b$, where $b < 1$, *a* is a constant. This means that with growth of mass the fraction of energy required to compensate gravitational force will be *increasing*.

We can write for metabolic powers $E_{Mg}$ and $E_{mg}$ required to compensate *gravitational* force for birds with mass *M* and *m* the following.

$$E_{Mg} = E_{mg} \frac{M}{m} \quad (6)$$

On the other hand, the relationship between *metabolic* powers $E_{Mv}$ and $E_{mv}$ for species with masses *M* and *m*, without accounting for allometric exponent for speed, is

$$E_{Mv} = E_{mv} \left(\frac{M}{m}\right)^{b_f} \quad (7)$$

The rationale behind not including the allometric exponent for velocity in (7) is that we want to find the fraction of energy used to support velocity, with which the allometric exponent for speed is associated. If we also include it into (7), that is if we use the total allometric exponent, that will mean that we included the associated part of energy used for velocity. However, to calculate the velocity exponent, we obviously need to account for energy supporting *velocity only*, extracting this part from the total energy expenditures, while associating the rest of energy with the allometric exponent $b_f$.

Now let us find the ratio of metabolic energies needed for counteracting gravity, and of energy supporting velocity, dividing equation (6) by equation (7).

$$\frac{E_{Mg}}{E_{Mv}} = \frac{E_{mg}}{E_{mv}} \left(\frac{M}{m}\right)^{1-b_f} \quad (8)$$

Therefore, once we know the ratio $r_{gv} = E_{mg}/E_{mv}$ for mass $m$, we can find the same ratio $R_{gv} = E_{mg}/E_{mv}$ for any mass $M$, so that

$$R_{gv} = r_{gv}\left(\frac{M}{m}\right)^{1-b_f} \quad (9)$$

Mass $m$ will be our reference point, with which we will associate a particular fraction of energy used to counteract gravity. Of course, such a reference point can be *any* mass, including the biggest one.

Just to be certain, we consider power $E_{gv}$ used for counteraction of gravity ($E_g$) and providing speed ($E_v$), so that $E_{gv} = E_g + E_v$.

The idea of the method, applied to *steady flights*, is to hypothetically transform all the power $E_{gv}$ into velocity (in general, to other meaningful parameter), as if a bird flies *without counteracting the gravity*. The rest of the total metabolic power supports allometric scaling for "biomechanical" part (the exponents $b_l$ and $b_{sk}$ accordingly in Eqn. (4), or the exponent $b_f$ in our notations).

Now, for the principal consideration. To calculate the allometric exponent, we will translate the metabolic power used for compensation of gravity into the hypothetical speed increase of the bird, thus obtaining value of such a hypothetical velocity $V_h$. In other words, we will find a speed, which the bird would develop if it utilized the already used energy for developing velocity, *plus* the energy to compensate the gravity. This is an important consideration. How can one translate energy increase into speed? It was theoretically shown in (Shestopaloff, 2016c, 2024a) that the spent energy is *proportional* to speed (Eq. (13) in the last work). The same result was earlier obtained experimentally (Taylor et al. 1970; Schmidt-Nielsen 1984). So, we assume proportionality between the additional energy and speed increase. Then, one can write:

$$V_h = V_r \left[ r_{gv} \left( \frac{M}{m} \right)^{1-b_f} + 1 \right] \quad (10)$$

where $V_r$ is the real observed velocity of a bird.

Once we calculate $V_h$, we can find allometric exponent, assuming we know $r_{gv}$. In fact, we will work the other way around, since it is the value of $r_{gv}$, which we want to find, while we know the allometric exponent $b_{SF} = 0.802$ for steady flights, and the value of $b_{ShF} = 0.874$ for short flights, corresponding to maximal metabolic rate.

## 3. Results

### 3.1. Calculating metabolic allometric exponent for a steady flight

Here, we apply the proposed method to particular data. Table 1 shows the data set for 17 birds, taken from (Shestopaloff, 2024a), Table 6D in Appendix D.3, which will be used for calculations. The birds were chosen to provide more uniform distribution of their mass over the range. Our goal is to find the ratio of metabolic powers used for compensation of gravity and supporting steady flights.

Table 1. Data set used for finding allometric exponent for hypothetical velocity $V_h$.

| Name | Mass, kg | Vel. km/h | $V_h$, km/h |
|---|---|---|---|
| Quail | 0.07 | 24 | 30.484 |
| Guinea_fowl | 0.7 | 35 | 56.415 |
| Barn_Owl | 0.3 | 80 | 116.234 |
| Budgerigar | 0.03 | 42 | 50.4 |
| Goose | 1.5 | 90 | 162.178 |
| Heron | 1.5 | 64 | 115.326 |
| Kingfisher | 0.01 | 40 | 45.416 |
| Macaw | 0.9 | 24 | 40.055 |

*Yuri K. Shestopaloff*

| | | | |
|---|---|---|---|
| Magpie | 0.2 | 32 | 44.550 |
| Pelican | 2.7 | 65 | 129.224 |
| Puffin | 0.368 | 88 | 130.855 |
| Robin | 0.016 | 29 | 33.639 |
| Sparrow | 0.0134 | 40 | 46.009 |
| Vulture | 0.85 | 48 | 79.465 |
| Falcon | 0.7 | 90 | 145.068 |
| Eagle | 5 | 50 | 111.483 |
| jackdaw | 0.25 | 24 | 34.1890 |

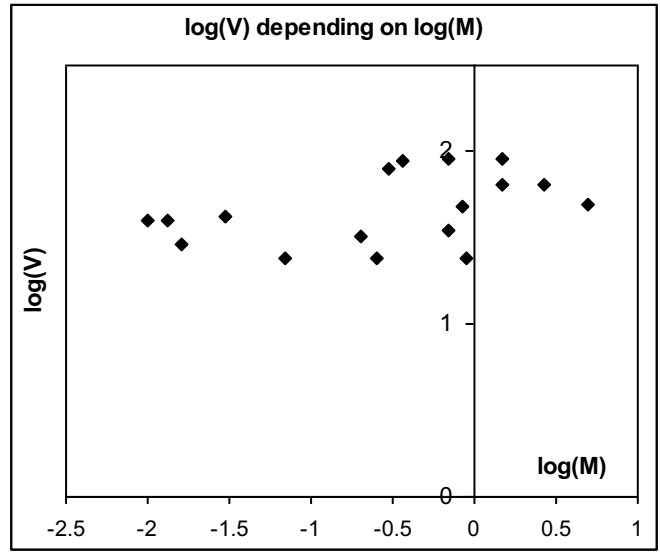

Fig. 1. Dependence of logarithm of flight velocity on the logarithm of birds' mass. $n = 17$.

Fig. 1 shows birds' velocities and their mass in logarithmic coordinates for 17 birds. The approach is as follows. Using (10), we calculate hypothetical velocity $V_h$, for several sequential values of ratio $r_{gv}$, for a reference mass 30 g. Then, we find the allometric exponent $b_v$ in logarithmic coordinates $log(M)$ and $log(V_h)$ for each value of this ratio, and then the total allometric exponent $b_x$ as $b_x = b_f + b_v = 0.645 + b_v$. Fig. 2 shows dependence of the calculated total allometric exponent $b_x$ from the ratio $r_{gv}$.

*Yuri K. Shestopaloff*

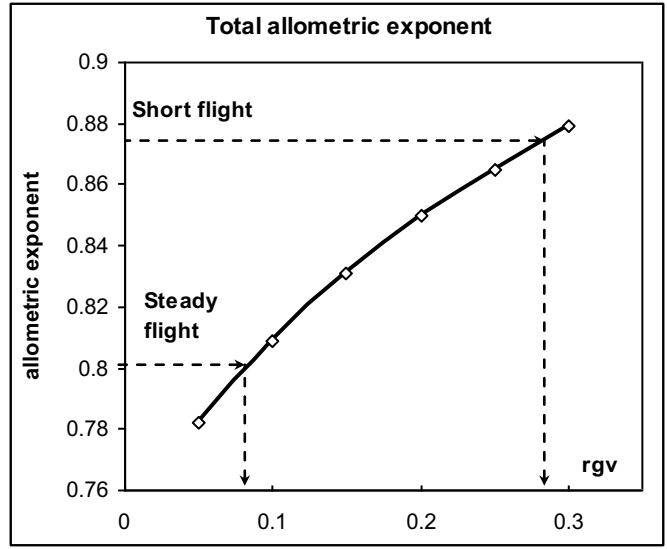

Fig. 2. Dependence of the total allometric exponent $b_x$ from the ratio $r_{gv}$ of metabolic powers used to compensate the force of gravity and supporting speed during steady and short flights of 30 g bird. Dashed lines project experimental values of allometric exponents to the axis $r_{gv}$.

Then, knowing the experimental value of the total allometric exponent $b_x = 0.802$ for steady flights, we find from the graph on Fig. 2 the ratio $r_{gv} \approx 0.08$ (for the bird with mass $m$=30 g). This means that the fraction of metabolic power used to counteract the gravity by such a small bird is about 8% of the metabolic power required to support speed during the steady flight. Certainly, using Eqn. 9, we can recalculate this ratio for any arbitrary mass $M$. For instance, for a bird with mass 20 kg we find $R_{gv} = r_{gv}(M/m)^{1-b_f} = 0.08(20/0.03)^{0.355} \approx 0.805$. In other words, the heavier the bird, the greater is the share of energy required to compensate the force of gravity, so that the monotonic increase of this share eventually caps the maximum mass of flying birds. Based on our numerical example, this share apparently should not noticeably exceed 80% of energy expenditures required to support speed during steady flight, since the chosen 20 *kg* value is close to the mass of the biggest known flying bird.

Dependence of ratio $R_{gv}$ on the birds' mass is shown on Fig. 3 (solid line). We can see that with increase of mass the fraction of power required to compensate gravity *increases*.

*Yuri K. Shestopaloff*

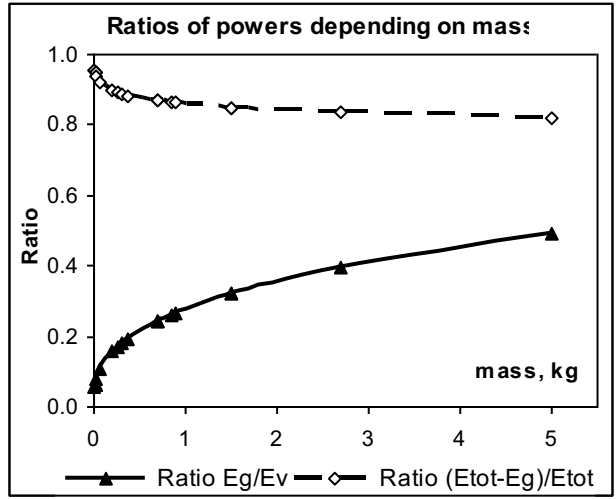

Fig. 3. Ratio of powers depending on mass. Solid line - $R_{gv}$ (ratio of power to compensate force of gravity to power required to support speed). Dashed line - ratio of total power $E_{tot}$ to support flight minus power to compensate gravity, to $E_{tot}$.

## 3.2. Energy expenditures in short flights

Similarly, we can find the fraction of metabolic power (in shares of power used to support speed during a steady flight), which is used for compensation the force of gravity, acceleration and acquiring a potential energy during ascending take-offs. As it follows from the graph on Fig. 2, for the allometric exponent $b_{ShF} = 0.874$, we have $r \approx 0.28$ for a bird of 30 g mass. For a bird of 20 kg mass, Eqn. 9 accordingly produces the ratio

$$R = r(M/m)^{1-b_f} = 0.28(20/0.03)^{0.355} \approx 2.8$$

We do not use index '$gv$', since this time the ratio accounts not only for the power needed to compensate force of gravity, but also for the power required for acceleration and to acquire potential energy during true take-offs.

Given the obtained number, it is very unlikely to see the same agility in a 20 kg bird, as in a 30 g avian creature, that is neither fast acceleration in take-offs with positive angle, nor expect quick vertical ascend. Indeed, heavy birds like geese, loonies ascend slowly at take-offs.

Previously we estimated in subsection 2.4 that during short flights acceleration and acquisition of potential energy during ascends require metabolic powers of about the same order of magnitude. Compensation of force of gravity requires an analogous magnitude of a metabolic power, as it follows from the above estimations. When mass of birds increases, they cannot spend metabolic power in the same proportions, because of the increase of energy de-

*Yuri K. Shestopaloff*

mands to compensate the force of gravity, as it was shown by numerical examples and by a graph in Fig. 3. So, they have to compensate such a required redistribution of energy in favor of compensation the force of gravity by reducing their acceleration characteristics, and by a slower vertical ascend to acquire potential energy.

Fig. 3 shows this consideration in a graphical form. We can see that the ratio $(E_{tot} - E_g)/E_{tot}$ monotonically decreases, where $E_{tot}$ is the power compensating gravity, supporting acceleration, speed and increase of potential energy during the flight. Such a decrease happens because the power needed for gravity compensation increases with the growth of mass. So, in relative terms, the greater is mass of a bird, the relatively less power is available to support speed, acceleration and the pace of vertical ascent.

### 3.3. Calculating the basal metabolic rate

Now we can calculate the basal metabolic rate. We already quoted in subsection 2.3 that "maximum flight metabolic rates generally exceed BMR by 8 to 14-fold for flying birds", while in (Nudds and Bryant, 2000) short flight energy expenditures for zebra finches were found of 27.8 times of BMR. Thus, in this regard, birds can be considered as athletic animals.

We will follow the same method for calculating the basal metabolic rate as in (Shestopaloff, 2016c, 2024a). The idea is that the "mechanical" part of the total allometric exponent (the one related to wing span and skeleton mass scaling) remains the same for the basal and maximal metabolic rate, which is a reasonable assumption. So, the difference is due to velocity part. And, since the maximal metabolic rate is much greater than the basal metabolic rate, we can adjust only the allometric exponent for velocity, that is to use the value of

$$(1/k)b_v \quad (11)$$

where $k$ defines how many times the maximal metabolic rate exceeds the basal metabolic rate. The value $b_v$ depends on the set of birds one uses for evaluation. For our restricted set of velocities from Table 1 we have $b_v \approx 0.104$. For a greater set from (Shestopaloff, 2024) the value $b_v \approx 0.126 \pm 0.04$ was obtained.

Using more conservative value of $k=15$, we find the basal metabolic rate, BMR, as follows.

$b_{x,BMR} = (1 - b_l)b_{sk} + (1/k)b_v = 0.645 + (1/15) \times 0.104 = 0.652$. For the aforementioned value $b_v \approx 0.126$, we accordingly obtain $b_{x,BMR} = 0.6534$.

Previously, we mentioned the range of values $b_{BMR} = 0.652$ and $b_{BMR} = 0.662$ for birds, reported in the literature. Both our obtained values for BMR neatly fall into this range. So, we may conclude that the obtained by our indirect method values of allometric exponents for BMR correspond to experimental data.

## 4. Discussion

At the beginning of this article we set the main goal of the study introduction of one broader concept that the metabolic allometric effect in living organisms is a consequence of natural selection guided by optimization of distribution of shared resources between the species of a food chain existing in a certain habitat. This optimization results in such an arrangement that bigger animals have some competitive advantage (say, a somewhat greater speed) sufficient to secure enough resources for sustainable reproduction of this species, but not too strong advantage to jeopardize existence of species they may prey on or which consume the same type of resources.

It is from this general concept a more particular method was derived for finding composition of energy expenditures supporting distinct organismal functions. This particular method was applied to metabolism of birds, and, indeed, we were able to find how much metabolic power is needed to compensate the force of gravity as a fraction of the power required to support speed during steady flight. We also showed that the metabolic power needed to compensate the force of gravity is an important factor, whose share increases with mass, in the overall metabolism of flying birds. We were also able to find how much relative metabolic power is required for acceleration and acquiring potential energy during short flights. It turned out that all these three components require about the same order of magnitude of metabolic power for a bird with mass of 30 g. Since the short flights exert the maximal physical workload on birds, therefore these three power components, together with the power needed to support speed, characterize the maximal energetic capabilities of birds.

It was found that the power needed to compensate the force of gravity during the flight increases in relative terms with the increase of mass of flying birds, so that available power has to be redistributed when the mass of flying birds increases. In this redistribution, less power is available for acceleration and faster acquisition of potential energy for ascending take-offs and flights with upward vertical component. What is important, the proposed method allows not only understanding and explaining these effects *qualitatively*, but also to *quantify* such characteristics depending on organismal parameters, such as mass, speed, metabolic power.

It is also important to note that the proposed method allows to quantify not only metabolic characteristics related to flight of birds, but also to find an allometric exponent for the basal metabolic rate (BMR). This fact positions the method as a rather universal tool for studies of animal metabolism in general, not only in birds. This advantageous feature of the method is rather the expression of universality of the *general concept* this method is based upon, while its particular application to birds' metabolism clearly demonstrated its high efficiency for solving difficult problems of metabolic studies in animals.

*Future development*

Effectively, the proposed approach, especially when integrated with the methods and ideas resulting from the concept that the metabolic allometric scaling effect is due to optimization of distribution of habitat's resources between the species of a food chain, existing in this habitat, represents rather a new ecological framework, within which new developments can evolve. Such, it would be useful to conduct experimental studies confirming results, theoretically obtained here. In this, the proposed approach could provide efficient guidance for experimental design and processing of experimental data.

The interesting effects discovered in this study certainly should be studied further both theoretically and experimentally. First it relates to more advanced validation of the proposed method applying it to more theoretical and practical problems. Secondly, it would be interesting to further explore the principal restriction on maximum mass of flying animals. This limitation may have very interesting and revealing implications and generalizations, and maybe not with regard to metabolism only.

## 5. Acknowledgements

The author thanks Dr. K.Y. Shestopaloff for the suggestions improving article's composition, editorial advice, and help with the literature.

*Yuri K. Shestopaloff*

*Yuri K. Shestopaloff*